\documentclass[graybox]{svmult}
\usepackage{type1cm}   
\usepackage{makeidx}   
\usepackage{graphicx}    
\usepackage{multicol}   
\usepackage[bottom]{footmisc}
\usepackage{newtxtext} 
\usepackage[varvw]{newtxmath} 
\usepackage{pgfplots}
\usepackage{multirow}
\usepackage{cite}
\usepackage[normalem]{ulem}
\usepackage{siunitx}

\pgfplotsset{compat=1.16}

\makeindex

\title*{Effect of crystallographic twins on the elastoplastic response of polycrystals}

\author{Lucas Monteiro Fernandes, Philipp Rieder, Matthias Neumann, Aude Mulard, Henry Proudhon, Volker Schmidt, Fran\c{c}ois Willot}
\authorrunning{Monteiro Fernandes et al.}
\institute{Lucas Monteiro Fernandes and Fran\c{c}ois Willot \at Mines Paris PSL, Center for Mathematical
Morphology, 35 rue Saint Honor\'e 77305 Fontainebleau, France \\ 
\email{lucas.monteiro\_fernandes@minesparis.psl.eu; francois.willot@minesparis.psl.eu} 
\and Philipp Rieder, Matthias Neumann and Volker Schmidt \at Ulm University, Institute for Stochastics, Helmholtzstr. 18 89069 Ulm, Germany \\
\email{philipp.rieder@uni-ulm.de ; matthias.neumann@uni-ulm.de ; volker.schmidt@uni-ulm.de}
\and Aude Mulard and Henry Proudhon \at Mines Paris PSL, Center for Materials Science,
63-65 Rue Henri-Auguste Desbru\`eres 91100 Corbeil-Essonnes, France \\ \email{aude.mulard@minesparis.psl.eu; henry.proudhon@minesparis.psl.eu} }

\begin{document}

\maketitle

\abstract{We investigate the influence of crystallographic twins on the elastoplastic response of $\gamma$-TiAl
intermetallics via full-field FFT-based computations. 
We first introduce a hierarchical stochastic model, which is used to simulate synthetic polycrystalline microstructures
containing twin grains with certain morphologies, and apply it to generate representative volume elements.
Second, we develop a Fourier-based method with regularization for solving the effective and local mechanical response of
polycrystalline media using the M\'eric-Cailletaud crystal plasticity constitutive law.
Numerical results show that, across configurations of twinning, the corresponding average effective response is similar.
Although differences were quantified, the effect of twins on the yield stress is negligible in practice (less than $1\%$).}

\section{Introduction}
Crystallographic twins are adjacent grains whose crystalline lattices are symmetric in a sense that they share
the points lying on their joint grain boundary. They may originate during certain thermodynamic conditions of
solidification, during lattice symmetry changes at the solid state, or during the response to shear stress.
Several authors have proposed to model twinning as a mechanism of plastic straining, generally via a crystal
plasticity framework. 
Clausen et al.~\cite{clausen_reorientation_2008} studied twinning in magnesium alloys and accurately modeled twin
activation and the subsequent textural evolution. 
They notably discussed stress relaxation mechanisms due to twinning.
Juan et al.~\cite{juan2014double} proposed a double-inclusion elastoplastic self-consistent scheme to model polycrystals
in twinning relations. This approach allows one to take into account relevant phenomena such as the initial stresses
experienced by
grains with respect to the stress state of their twin domain.
The model is based on the assumption of ellipsoidal inclusions embedded in a homogeneous medium,
common in homogenization schemes.
Local effects that arise from the interaction between grain boundary geometries and crystalline orientations are
not explicitly represented.
Zhang~\textit{et al}~\cite{zhang_simulation_2023} proposed a coupled FEM-FFT model to simulate dislocation slip
and twinning propagation during the compression of magnesium micro-pillars, and compared the predictions of the
numerical model to experimental data.
Numerical predictions show that for a 2D polycrystalline microstructure, in a scenario where the grains are in
a twin relation, the ultimate stresses are lower compared to a scenario where no twins are modeled. Numerical
results and analytical estimates obtained from self-consistent homogenization theories
also indicate a stiffer response for polycrystals without twins~\cite{rieder2023}.

The present work is concerned with the elastoplastic response of twinned structures, derived from a stochastic
3D microstructure model.
In particular, we investigate by numerical tools how twin grains influence the effective response of
polycrystals at a macroscopic scale.
The microstructure model, presented in Section~\ref{sec:stochastic}, is used to generate grain architectures
of polycrystaline materials and introduces the grain-wise crystallographic orientations
that model twinning relations.
We consider a crystal plasticity law based on small strains explicitly integrated locally, while taking into
account not only dislocation slip systems but also pseudo-slip systems modeling micro-twinning. 
The constitutive law coupled to a Green operator-based FFT solver used to solve the mechanical problem
are described in Sections~\ref{sec:mclaw} and~\ref{sec:regularization}. 
The obtained results are presented and discussed in Section~\ref{sec:results}.
Section~\ref{sec:conclusion} concludes.

\section{Stochastic 3D modeling of polycrystalline grain architectures}\label{sec:stochastic}
In this section we outline the stochastic 3D microstructure model whose realizations serve as geometric
input for the numerical simulations described in Section \ref{sec:results}. The model is realized within a
cubic sampling window $W \subset \mathbb{R}^3$ and implemented with periodic boundary conditions to avoid
edge effects. For a more comprehensive description of this model, we refer to \cite{rieder2023}.

\subsection{Space tessellation model}
In particular,
in the present work, a Voronoi tessellation is employed, which subdivides the Euclidean space $\mathbb{R}^3$ into
subsets with disjoint interiors, 
called cells in stochastic geometry~\cite{chiu.2013} and grains in the context of crystallography.
The $n$-th grain $G_n\subset\mathbb{R}^3$ of the tessellation is given by the set of those points $x\in\mathbb{R}^3$,
which are not further away (with respect to the Euclidean distance) from some generator point $s_n$ than
from all other generator points $S_m$ ($m\neq n$):
\begin{align}
    G_n=\bigl\{ x \in \mathbb{R}^3 : |x-s_n | \leq |x-S_m |  \text{ for all } m\neq n   \bigr\}, 
\end{align}
where $| \cdot |$ denotes the Euclidean distance. Note that the stochastic nature of random Voronoi
tessellations arises from the randomness of the underlying point process  of generator points. 
An inherent property of Voronoi tessellations is that the grains are convex polyhedrons, in particular
their boundaries are given by planar facets. Furthermore, a Voronoi tessellation is uniquely characterized
by its generating point pattern. In this study, the generating point pattern
 $\{s_n\}$ in $\mathbb{R}^3$ is sampled from a Mat\'ern-hardcore point process with intensity $\lambda>0$
 and hardcore radius $r\geq 0$. The Mat\'ern-hardcore process is obtained by a thinning of a homogeneous Poisson
 process~\cite{last.2017}, such that the minimum pairwise distance between two points of the thinned point pattern is
 larger than $r$. This choice of a point-process model allows us to control the grain size of the corresponding Voronoi
 tessellation via the hardcore radius $r$. For a formal definition of  Mat\'ern-hardcore point processes, we refer to~\cite{chiu.2013}.
Once a Mat\'ern-Voronoi tessellation has been generated,
one must define the crystallographic orientation within each grain which will be addressed in the following section.

\subsection{Modeling of crystallographic orientations and twinning effects}
\begin{figure}[t]
    \centering\includegraphics{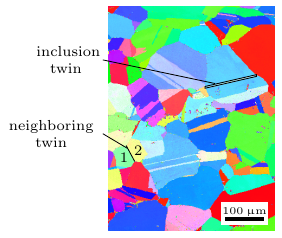}
    \caption{Electron backscatter diffraction (EBSD) scan of a $\gamma$-TiAl sample. Pairs of twin grains
    with different morphologies can be observed.}
    \label{fig:ebsd}
\end{figure}

A twin relation between two grains depends not only on the
grains respective crystallographic orientations, but also on the spatial orientation of their joint grain boundary. 
Crystallographic twins are pairs of neighboring grains, which are in a certain relationship to each other.
On the one hand, the crystallographic misorientation between such neighboring grains, as well as the spatial
orientation of their joint grain boundary
have to fulfill two main criteria: the misorientation must coincide with a predefined angle and the
crystallographic habit plane
has to be aligned parallel to the joint grain boundary of the neighboring grains in a twin relation.
In general, these properties of twin relations depend on the type of crystal symmetry. For the tetragonal
crystal symmetry of $\gamma$-TiAl, considered in this paper, the misorientation angle $\theta$ has to be equal
to $2 \, \text{tan}^{-1}(a/c)\approx 69.79^\circ$, where $a$ and $c$ denote the lattice constants. Furthermore,
the habit plane is   either the $(011)$ or $(101)$ plane of the tetragonal lattice. The model used in the present
study incorporates two different types of twins, which we treat separately. 
These two types of twins have been observed on EBSD maps of $\gamma$-TiAl polycrystalline aggregates, see
Figure~\ref{fig:ebsd}.
The first type, referred to in this study as ``neighboring twins'', are modeled by randomly choosing contiguous pairs
of grains.
For each of these pairs, the crystallographic orientations of the adjacent grains are determined such that they are in a twin relation.
For the second type of twins, which are labeled as ``inclusion twins'', a crystallographic orientation is attached to randomly chosen grains
from the initial tessellation, the so-called parent grains. For each of them, depending on their crystallographic
orientation, another grain, the child grain, is inserted in the interior of the parent grain,
such that they are in a twin relation, as specified hereafter.

\subsubsection{Neighboring twins}
To determine the number of neighboring twins within the sampling window $W\subset\mathbb{R}^3$, we realize a Poisson-distributed
random variable $N_{\text{neigh}}$ with expectation equal to $p_\text{neigh}\cdot N_W/2$, where $N_W\in\{1,2,\ldots\}$ denotes
the number of grains contained in $W$, and $p_\text{neigh}\in[0,1]$ a neighboring twins parameter. Note that the realization of
$N_\text{neigh}$ is rejected until $N_\text{neigh}\leq N_W/2$. 
The following procedure randomly generates pairs of twin grains, by assigning them with a crystallographic orientation.
The procedure is repeated until  $N_\text{neigh}$ pairs  are created, or no grains with free neighbors are left. 
It is important to note that once a grain is chosen and assigned a crystallographic orientation, it cannot be chosen a second time.
Consequently, the structure contains no so-called twin related domains, where twin grains form a ``chain'' with several grains.

In order to model the crystallographic orientations of neighboring twins, we proceed as follows. At first,  a 
grain $G_k$
and one of its neighbours, denoted by  $G_\mathbf{l}l$, are chosen at random.  Let
$\partial G_{k,\mathbf{l}l}=\partial G_{\mathbf{l}l,k}=G_k \cap G_\mathbf{l}l$ denote the joint grain boundary of $G_k$
and $G_\mathbf{l}l$. Recall that the Voronoi tessellations form polyhedral grains, which implies that $\partial G_{k,\mathbf{l}l}$
is a two-dimensional facet. 
Furthermore, the crystallographic orientation $O_k$ of grain $G_k$ is sampled from a uniform distribution on the space of
rotations SO$_3$, under the condition that one of its habit planes aligns with the joint grain boundary
$\partial G_{k,\mathbf{l}l}$. Subsequently, the orientation of $O_\mathbf{l}l $ of $G_\mathbf{l}l$ is determined
by rotating  $O_k$ by the misorientation angle $\theta$ around $[100]$ or $[010]$, respectively.

\subsubsection{Inclusion twins}
So far, only those grains involved in neighboring twin relations are assigned with a crystallographic orientation, in general
some grains are left without an orientation. These remaining orientations are sampled from a uniform distribution on the space of rotations SO$_3$.

The generation of inclusion twins is controlled by a probability $p_{\text{incl}}\in[0,1]$ such that a grain $G$ is equipped
with an inclusion $I\subset G$, which splits $G$ into two parts. If $G$ is selected to get an inclusion, the following procedure
is executed, otherwise the algorithm continues with the next grain.

First, the crystallographic orientation $O_I$ of $I$ is determined by rotating $O$, the orientation of $G$, by the
misorientation angle $\theta$ around a randomly chosen $[100]$ or $[010]$ direction.
Note that the choice of  the rotation axis uniquely determines the spatial orientation of the habit plane (up to translations).

Second, we describe how the inclusion twin $I$ is defined, namely  as dilated
plane, parallel to the habit plane.
Let $\partial G$ denote the grain boundary of $G$ and $x\in\partial G$ a  point on it chosen at random. 
Furthermore, we define $\mathcal{H}$ as plane, that is parallel to the habit plane and contains $x$.  Additionally, we
introduce $\mathcal{H}_\delta$ as $\mathcal{H}$, which is dilated to a thickness of $\delta>0$ and define $I$ as intersection
of $G$ and $\mathcal{H}_\delta$.
The value of $\delta$ is sampled at random on 
$[\delta_-\hspace{-3pt}\cdot \! \text{ved}(G),\delta_+\hspace{-3pt}\cdot \! \text{ved}(G)]\subset[0,\text{ved}(G)]$, 
where $\text{ved}(G)$ denotes the volume equiv\-a\-lent diameter of $G$ with $0\leq\delta_-\leq\delta_+\leq1$, the bounds
of the relative thickness of the inclusion.

For a visualization of the generation of inclusion twins, we refer to \cite{rieder2023}.

\subsection{Specification of model parameters}
We draw a total of 15 realizations from the stochastic model described above

for each of the following twinning configurations:
(i) neighboring twins ($p_{\text{neigh}}=1, p_{\text{incl}}=0$),
(ii) inclusion twins ($p_{\text{neigh}}=0, p_{\text{incl}}=1$)
and (iii) polycrystals without twins ($p_{\text{neigh}}=p_{\text{incl}}=0$).
Note that for neighboring twins,
although $p_\text{neigh}$ is equal to 1, not every grain necessarily participates in a twin relationship.
The sampling window $W$  contains $256^3$ voxels. The parameters of the Mat\'ern-hardcore point process are given by a hardcore
radius $r$ of 5 voxels and an intensity $\lambda=1920/(256 ~\mathrm{voxels}^3)= \approx1.14\cdot10^{-4} / \mathrm{voxels}^3$,
which corresponds to 1920 expected grains within the sampling window $W$.
To reduce edge effects, which may influence the results of the mechanical simulations described in Section \ref{sec:results}, 
realizations of the stochastic model are generated with periodic boundary conditions. 

\begin{figure}
\begin{center}
\begin{tabular}{ccc}
\includegraphics[trim={2.75cm 2.75cm 2.75cm 2.75cm},clip,width=0.3\linewidth]{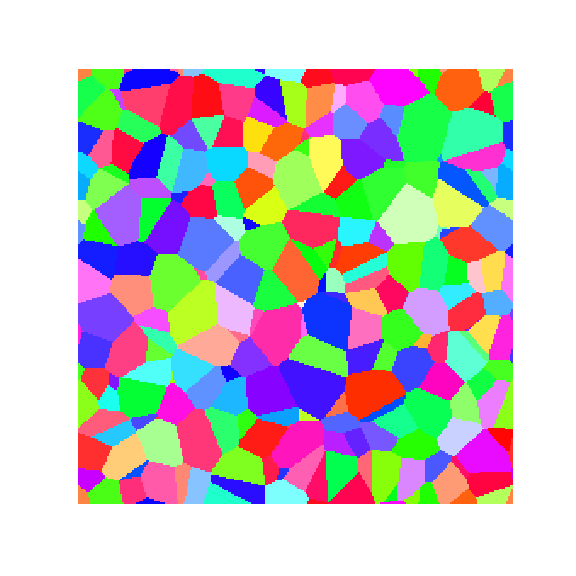} & 
\includegraphics[trim={2.75cm 2.75cm 2.75cm 2.75cm},clip,width=0.3\linewidth]{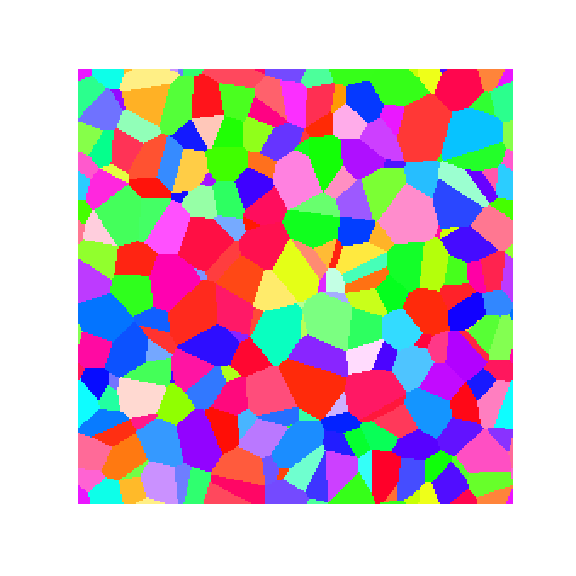} &
\includegraphics[trim={2.75cm 2.75cm 2.75cm 2.75cm},clip,width=0.3\linewidth]{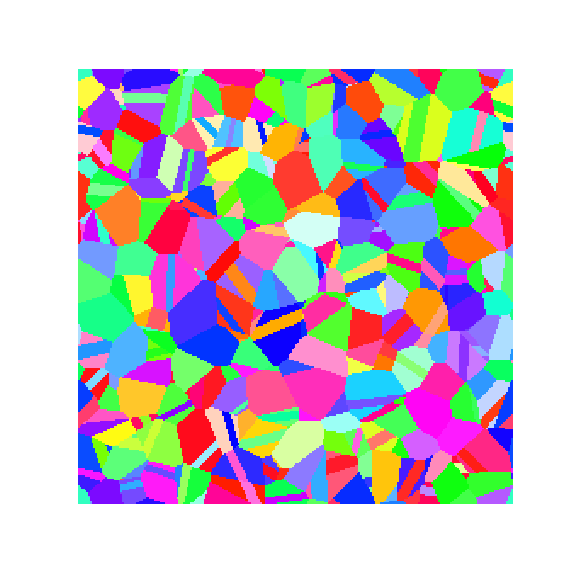} 
\\ (a) & (b) & (c)
\end{tabular}
\caption{Examples of inverse pole figure (IPF) colored 2D crops of the generated 3D microstructures:
a) Realization without twins.
b) The same realization following the insertion of local correlations between orientations for modeling neighboring twins.
c) The same realization following the insertion of inclusion twins, with both morphological and crystallographic alteration.}
\label{fig:realizations}
\end{center}
\end{figure}

\section{M\'eric-Cailletaud crystal plasticity}\label{sec:mclaw}
The single-crystal plasticity model presented in this article is inspired by the one described by M\'eric and Cailletaud
in~\cite{meric_single_1991} and experimentally fitted for a  $\gamma$-TiAL sample by Mulard in~\cite{mulard_micro-mechanical_2018},
from which the relevant parameters were used in the following. In the present study, we consider a crystalline material that possesses 12
slip systems relating to movement of dislocations and superdislocations and 4 pseudo-slip systems relating to micro-twinning mechanisms.
Assuming the validity of small strain theory, one can separate the strain tensor $\mathbf{\varepsilon}$ into elastic
$\mathbf{\varepsilon}^e$ and plastic $\mathbf{\varepsilon}^p$ contributions by
$\mathbf{\varepsilon}=\mathbf{\varepsilon}^e + \mathbf{\varepsilon}^p$.
$\mathbf{\varepsilon}^e$ is linked to the stress tensor $\mathbf{\sigma}$ via the Hooke's law with an anisotropic stiffness tensor $\mathbb{C}$, i.e.
$    \mathbf{\sigma}=\mathbb{C}:\mathbf{\varepsilon}^e$.
The plastic contribution on strain is determined by integrating the plastic strain rate $\dot{\mathbf{\varepsilon}}^v$, i.e.,
\begin{equation}
\dot{\mathbf{\varepsilon}}^v =  \left(1-\sum_{\beta=1}^{N_{\beta}}f^{\beta}\right)\sum_s^{N_s}\dot{\gamma}^s\mathbf{m}^s+
\sum_{\beta}^{N_{\beta}}\dot{\gamma}^{\beta}\mathbf{m}^{\beta}\,,
\end{equation}
where $N_{s}$ and $N_\beta$ denote respectively the number of dislocation and superdislocation $s$ slip systems and micro-twinning $\beta$
pseudo-slip systems. Further, $f^{\beta}$ denotes the volume fraction of micro-twinnning at the considered pseudo-slip system
$\beta$, $\dot{\gamma}^{s}$ and $\dot{\gamma}^{\beta}$ represent the slip rate and $\mathbf{m}^{s,\beta}$ is the Schmid tensor.
The tensor $\mathbf{m}^{s,\beta}$ compactly describes a slip system by condensing slip planes and directions into the tensor
\begin{equation}
      \mathbf{m}^{s,\beta}=\frac12\left(\mathbf{n}^{s,\beta}\otimes\mathbf{l}^{s,\beta}+\mathbf{l}^{s,\beta}\otimes\mathbf{n}^{s,\beta}\right),
\end{equation}
where $\otimes$ denotes the outer product. The driving force of slip is the resolved shear stress $\tau^{s,\beta}$, obtained by projecting
the stress tensor at both a slip plane and a slip direction. Doing so, we obtain
\begin{equation}
    \tau^{s,\beta}=\mathbf{\sigma}:\mathbf{m}^{s,\beta}.
\end{equation}
Increase in slip depends on whether the driving force has reached a critical value, called the yield stress $r^{s,\beta}$ for the slip system 
\begin{eqnarray}
r^s=\tau_0^{s}+Q_s\sum_{s'=1}^{N_s}q_{ss'}\left(1-\textnormal{e}^{-b_sv^{s'}}\right), \quad
r^{\beta}=\tau_0^{\beta}+Q_{\beta}\sum_{\beta'=1}^{N_{\beta}}q_{\beta \beta'}\left(1-\textnormal{e}^{-b_{\beta}v^{\beta'}}\right)
\end{eqnarray}
under consideration, where $\tau_0^{s,\beta}$ denotes the initial resistance to slip, $Q_{s,\beta}$ is the magnitude of hardening interaction,
$q_{ss',\beta \beta'}$ is the hardening interaction matrix. Here, $s'$ and $\beta'$ denote slip and pseudo-slip system indices respectively,
$b_{s,\beta}$ is the hardening saturation and $v^{s',\beta'}$ is the accumulated slip, which is obtained by integrating the absolute value of
the slip rate at each slip and pseudo-slip system: $\dot{v}^{s',\beta'}=|\dot{\gamma}^{s',\beta'}|$.
In order to determine the slip rates, not only the yield stresses are taken into account but also some viscous parameters $K_s$ and $n$ from
Norton's law. For dislocation slip systems, this reads as:
\vspace{-0.1cm}
\begin{equation}
    \dot{\gamma}^s=\left\langle\frac{|\tau^s|-r^s}{K_s}\right\rangle^n\textnormal{sign}(\tau^s).
\end{equation}
For micro-twinning pseudo-slip systems, the relation that exists between slip and the volume fraction of twins must be accounted for, namely
$\dot{\gamma}^{\beta}=\dot{f}^{\beta}/\sqrt{2}$.
Indeed, a certain volume becomes a micro-twin if the required shear displacement is a fraction of the interatomic distance, which is
determined by the geometry of the lattice and called twinning shear $\gamma^T$. For the material under consideration, $\gamma^T=1/\sqrt{2}$.
One then derives:
\begin{equation}
    \dot{f}^{\beta} = K_c\left(f_m - \sum_{\beta}f^{\beta} \right)\left\langle \frac{\tau^{\beta}-r^{\beta}}{K_t} \right\rangle^{n_t},
\end{equation}
where $K_c$ is the material-dependent magnitude of twinning and $f_m$ is a parameter modeling the maximum twinned volume fraction observed
experimentally in a grain. $K_t$ and $n_t$ are parameters from Norton's law in case of twinning pseudo-slip systems. It is important to
highlight that, contrarily to dislocation and superdislocation slip, twinning is unidirectional.
The formulation was implemented with an explicit discrete integration scheme (Runge-Kutta 4th order method) on a spectral solver based
on the Green's operator (commonly known as FFT solver).
As for elasticity, the following moduli have been measured experimentally~\cite{tanaka1996single} at $298$K:
\begin{equation}
c_{11}=183,\quad
c_{33}=183,\quad
c_{12}=c_{13}=74,\quad
c_{44}=105, \quad
c_{66}=78 \textnormal{ GPa}.
\end{equation}
In addition, we follow~\cite{mulard_micro-mechanical_2018} (Tab.~5.7, p.~79) and use the following single-crystal plasticity
parameters found by a coupled numerical and experimental identification:
\begin{subequations}
\begin{eqnarray}
K_s&=&K_t=  \SI{104}{\mega\pascal\second\tothe{1/n}} , \quad K_c=1.45, \quad n=n_t=6.42,  
\quad Q_s=\SI{9.57}{\mega\pascal},
\\ Q_{\beta}&=&\SI{10.5}{\mega\pascal} , \quad f_m=0.8, 
\quad \tau_0^{s,\mathcal{O}}=\tau_0^{\beta}=\SI{130}{\mega\pascal} ,\quad 
\tau_0^{s,\mathcal{S}}=\SI{200}{\mega\pascal},
\end{eqnarray}
\label{eq:plastic_params}
\end{subequations}
Here $\tau_0^{s,\mathcal{O}}$, $\tau_0^{s,\mathcal{S}}$ and $\tau_0^{\beta}$ refer to the initial resistance to slip respectively for
dislocation and superdislocation slip systems and for pseudo-slip systems associated to micro-twinning.

\begin{table}[t]
\begin{center}
\resizebox{\columnwidth}{!}{\begin{tabular}{c|c|c|c|c|c|c|c|c|c|c|c|c||c|c|c|c|c}
$s$               & 1  & 2 & 3                  & 4 & 5 & 6                   
       & 7 & 8 & 9                    & 10 & 11 & 12                 
  & $\beta$ & 1       & 2           & 3          & 4 
  \\ \hline
$\mathbf{n}^s$     & \multicolumn{3}{c|}{(11$\overline{1}$)}& \multicolumn{3}{c|}{($\overline{1}$ 11)} 
     & \multicolumn{3}{c|}{(1$\overline{1}$ 1)}& \multicolumn{3}{c||}{(111)}   
  & $\mathbf{n}^{\beta}$ & (11$\overline{1}$) & ($\overline{1}$ 11)    & (1$\overline{1}$ 1)   & (111) 
  \\ \hline
$\mathbf{l}^s$ & [011] & [101] & [1$\overline{1}$ 0]    & [0$\overline{1}$ 1] &
 [$\overline{1}$ 0$\overline{1}$] & [$\overline{1}$$\overline{1}$ 0] & [011] & 
 [10$\overline{1}$] & [011]      & [01$\overline{1}$] & [10$\overline{1}$] & [1$\overline{1}$ 0] 
  & $\mathbf{l}^{\beta}$ & [112]   & [$\overline{1}$ 1$\overline{2}$] & 
  [1$\overline{1}$$\overline{2}$] & [11$\overline{2}$]
  \\ \hline
type              & s & s & o                   & s & s & o                   
       & s & s & o                    & s & s & o                    
\end{tabular}
}
\end{center}
\caption{\label{tab:slipsystems}
Left: planes
$\mathbf{n}^{s}$ and directions $\mathbf{l}^{s}$
of dislocation slip systems of the $\gamma$-TiAl single-crystal; ``o'' and ``s'' denote respectively dislocations
and superdislocations.
Right: 
planes $\mathbf{n}^{\beta}$ and directions $\mathbf{l}^{\beta}$
of the pseudo-slip systems associated to the twinning of the $\gamma$-TiAl single-crystal. }
\end{table}

\section{Regularization by viscous hardening}\label{sec:regularization}
Wen the norm of the local stress tensor is
larger than $E\delta\overline{\varepsilon}$, where $E$ is the Young's modulus and $\delta\overline{\varepsilon}$ the 
strain increment, mechanical
equilibrium is not enforced in certain voxels. Algorithms such as Newton's method~\cite{besson2009non} can be used to deal 
with this problem, which is commonly encountered in the case of finite strains~\cite{ling2018reduced}.
In the present case, we make use of an
explicit scheme with a regularization technique inspired by viscous hardening, as a way
to obtain convergence even with large strain increments in the elastoplastic regime.
Namely, we introduce a parameter $\eta$ which instantaneously increases the yield
stresses for each slip and pseudo-slip system by a term proportional to the absolute value of the slip rate at the
considered slip and pseudo-slip system:
\begin{equation}\label{eq:regu}
r'^{s,\beta}=r^{s,\beta} + \eta|\dot{\gamma}^{s,\beta}|.
\end{equation}
This regularization term can be interpreted as viscosity that penalizes sudden and high variation in slip, and privileges
elastic deformation instead of plastic deformation. By regularizing, elastoplastic strain steps alternates with purely
elastic strain steps in the elastoplastic regime.
When $\eta$ takes values inside a certain range, the regularized solution varies little enough to be consistent. On the
one hand, if $\eta$ is too low, convergence is not obtained. On the other hand, if it is too high, purely elastic behavior
is recovered, which is not physically valid.
In the present work, we set $\eta=10^4$ MPa$\cdot$s.

\begin{figure}[t]
    \centering
    \begin{tikzpicture}
    \begin{axis}[
        xlabel={$\eta \, \text{(MPa}\cdot\text{s)}$},
        ylabel={ $ \text{Yield stress} \, \text{(MPa)}$},
        ymin=259, ymax=262, xmode=log, log basis x=10, grid=major, width=10cm, height=4cm      
    ]
    \addplot[blue,scatter] table {fig3.dat};
    \end{axis}
    \end{tikzpicture}
    \caption{Influence of the linear viscous hardening parameter $\eta$ on the yield stresses
    for the integration of the constitutive behavior and arbitrary imposed strain. It can assume
    a wide range of values provided it is high enough to avoid divergence of the simulation and
    low enough to prevent unrealistically high hardening.}
    \label{fig:enter-label}
\end{figure}
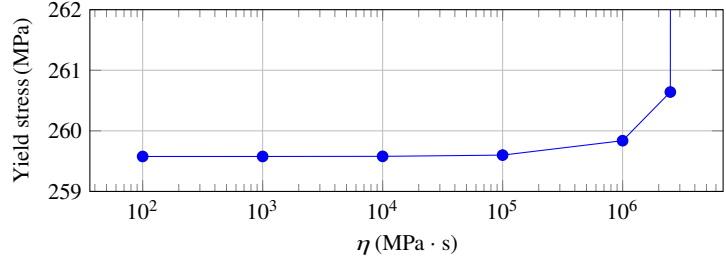

\section{Elastoplastic response of twinned polycrystals}\label{sec:results}
In order to compare the configurations of crystallographic twins, simulations were carried out for prescribed macroscopic uniaxial
shortening strain of $0.5~\%$.
Accordingly, we consider both the equivalent stress of the spatial average of the stress tensor field ${\langle \sigma \rangle}_{eq}$
and the spatial average of the equivalent stress field ${\langle \sigma_{eq} \rangle}$ as measures of the macroscopic response. Then,
realization-wise averages of these quantities can be computed for each configuration of twinning.

\begin{figure}[t]
\centering
\begin{tabular}{cc}
\pgfplotsset{width=0.45\linewidth, height=0.4\linewidth,compat=1.9}
\begin{tikzpicture}
\begin{axis}[
title= ,
xlabel={Uniaxial shortening strain $(\%)$},
ylabel={$\overline{{\langle \sigma \rangle}_{eq}}$ (MPa)},
legend style={at={(0.627,0)},anchor=south,nodes={scale=0.8, transform shape}},
xmin=0, xmax=0.5,
ymin=0, ymax=165,
]
\addplot[blue] table {fig4aa.dat};
\addplot[green] table {fig4ab.dat};
\addplot[red] table {fig4ac.dat};
\legend{
No twins,
Neighboring twins, 
Inclusion twins,
}
\end{axis}
\end{tikzpicture} &
\pgfplotsset{width=0.45\linewidth, height=0.4\linewidth,compat=1.9}
\begin{tikzpicture}
\begin{axis}[
title= ,
xlabel={Uniaxial shortening strain $(\%)$},
ylabel={$\overline{{\langle \sigma \rangle}_{eq}}$ (MPa)},
legend style={at={(0.627,0)},anchor=south,nodes={scale=0.8, transform shape}},
xmin=0.4, xmax=0.5,
ymin=158, ymax=163,
]
\addplot[blue] table {fig4aa.dat};
\addplot[green] table {fig4ab.dat};
\addplot[red] table {fig4ac.dat};
\legend{
No twins,
Neighboring twins, 
Inclusion twins,
}
\end{axis}
\end{tikzpicture}
\end{tabular}
\caption{
Von Mises stresses of the RVE-wise average of the stress tensors vs. strain curves of polycrystalline RVE realizations representing twin configurations.
Prescribed macroscopic uniaxial shortening strain along the $x$ axis. In the elastoplastic regime, small differences are observed between the three
configurations.}
\label{fig:stressstrain}
\end{figure}
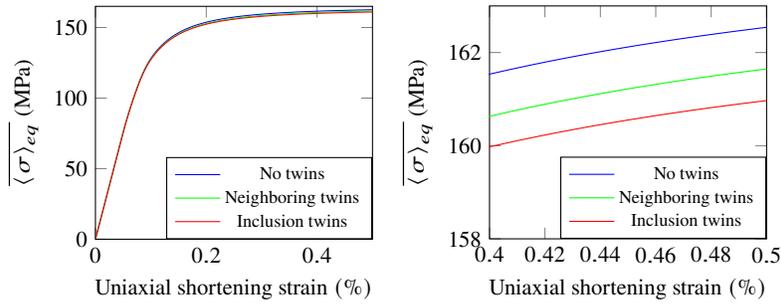

\begin{figure}
\begin{center}
\begin{tabular}{cccc}
\includegraphics[trim={2cm 1cm 2cm 1cm},clip,width=0.28\linewidth]{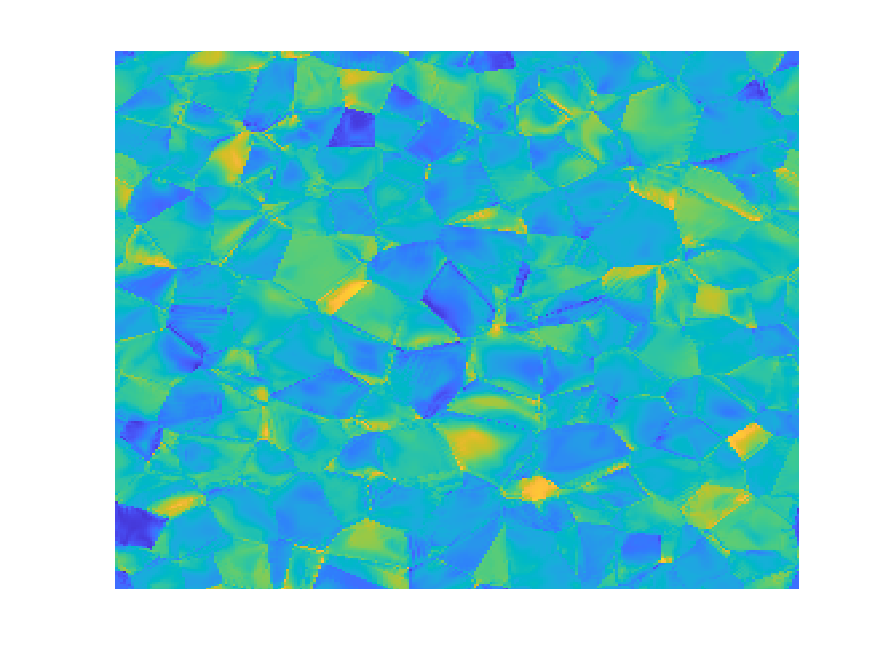} & 
\includegraphics[trim={2cm 1cm 2cm 1cm},clip,width=0.28\linewidth]{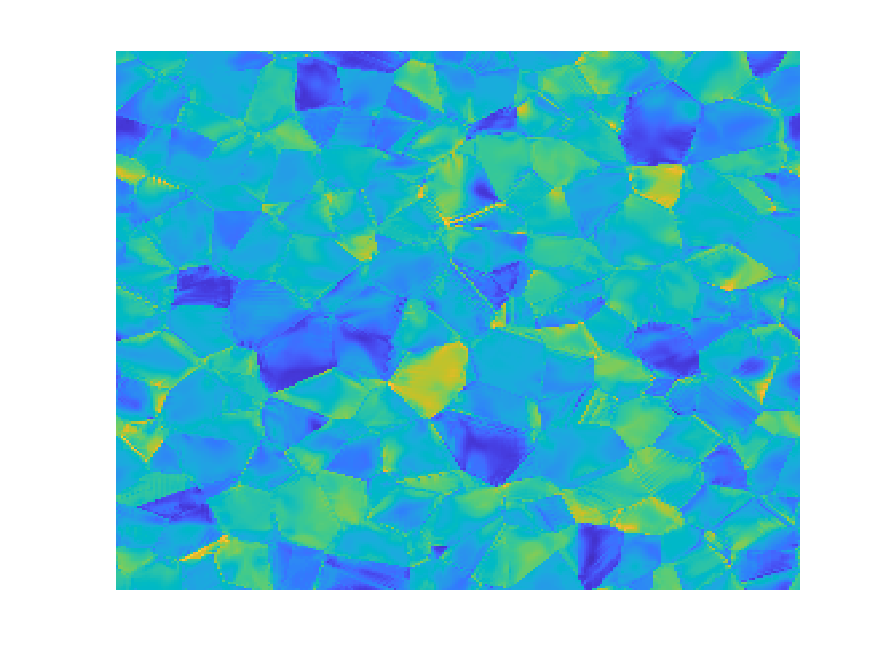} &
\includegraphics[trim={2cm 1cm 2cm 1cm},clip,width=0.28\linewidth]{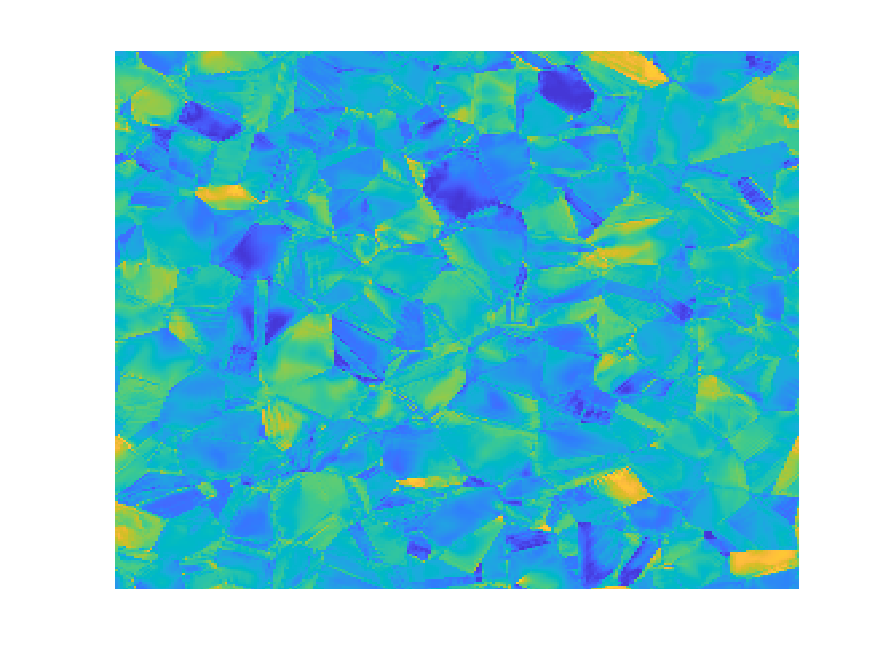} &
\includegraphics[trim={12cm 0.8cm 0cm 0cm},clip,width=0.073\linewidth]{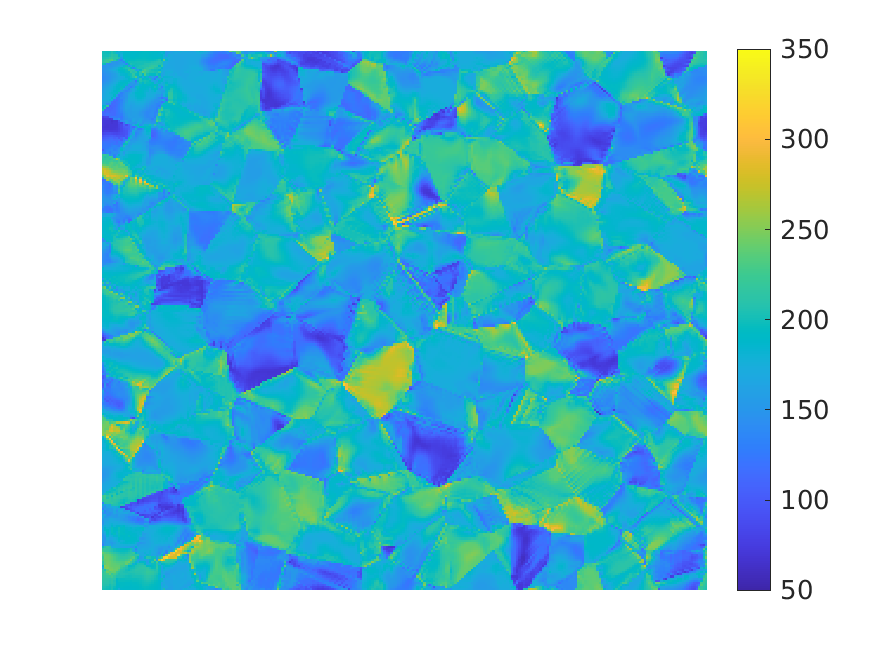} 
\\ (a) & (b) & (c)
\end{tabular}
\caption{Examples of Von Mises stress fields obtained with the FFT computations (Color map values in MPa),
corresponding to the images in Fig.~(\ref{fig:realizations}):
a) Realization without twins.
b) The same realization following the insertion of neighboring twins.
c) The same realization following the insertion of inclusion twins.
}
\label{fig:vmfields}
\end{center}
\end{figure}

For each configuration, fifteen realizations of the stochastic model described in Section~\ref{sec:stochastic} with a
size of $256^3$ voxels were simulated. Fig.~\ref{fig:stressstrain} shows averaged macroscopic stress-strain curves,
more precisely the realization-wise average of Von Mises equivalent stresses of the RVE-wise average stress tensors
as a function of the prescribed macroscopic strain at each strain step of the quasi-static simulation (1000 steps in total).
The observed average response is rather similar across the three configurations of twin grains.

\begin{table}[t]
\centering
\begin{tabular}{|c|c|c|c|c|}
\hline
    & Average nb. of grains & Average grain size & SD(\# grains) & SD(grain size) \\ \hline
Neighboring twins & 1932         & 8689   & 44       & 201 \\ \hline
Inclusion twins & 5589         & 3003   & 132       & 71 \\ \hline
No twins & 1932         & 8689   & 44       & 201 \\ \hline
\end{tabular}    \caption{\label{tab:grain_nb_and_size} 
Realization-wise averages and standard deviations (SD) of grain numbers and sizes for each configuration of twinning.
Unit of grain size: voxels
or $\mu\text{m}^3$.}
\end{table}

\begin{table}[t]
\centering
    \begin{tabular}{|cc|c|c|c|}
    \hline
    \multicolumn{2}{|c|}{}                            & Load along x      & Load along y     & Load along z      \\ \hline
    \multicolumn{1}{|c|}{\multirow{4}{*}{Neighboring twins}} & $\overline{{\langle \sigma \rangle}_{eq}}$  & 161.65  & 161.44 & 161.91  \\ \cline{2-5} 
    \multicolumn{1}{|c|}{}     & $\overline{\langle \sigma_{eq} \rangle}$  & 186.06  & 185.89 & 186.29  \\ \cline{2-5} 
    \multicolumn{1}{|c|}{}     & SD(${\langle \sigma \rangle}_{eq}$) & 1.36  & 1.28 & 1.01  \\ \cline{2-5} 
    \multicolumn{1}{|c|}{}    & SD(${\langle \sigma_{eq} \rangle}$) & 1.30  & 1.15 & 1.00 \\ \hline
    \multicolumn{1}{|c|}{\multirow{4}{*}{Inclusion twins}} & $\overline{{\langle \sigma \rangle}_{eq}}$  & 160.97  & 160.33 & 160.92  \\ \cline{2-5} 
    \multicolumn{1}{|c|}{}       & $\overline{\langle \sigma_{eq} \rangle}$  & 185.20   & 184.59 & 185.14  \\ \cline{2-5} 
    \multicolumn{1}{|c|}{}      & SD(${\langle \sigma \rangle}_{eq}$) & 0.78 & 1.22 & 0.67 \\ \cline{2-5} 
    \multicolumn{1}{|c|}{}        & SD(${\langle \sigma_{eq} \rangle}$) & 0.73 & 1.05 & 0.63 \\ \hline
    \multicolumn{1}{|c|}{\multirow{4}{*}{No twins}} & $\overline{{\langle \sigma \rangle}_{eq}}$  & 162.54  & 162.26 & 162.29  \\ \cline{2-5} 
    \multicolumn{1}{|c|}{}      & $\overline{\langle \sigma_{eq} \rangle}$  & 187.04  & 186.69 & 186.84  \\ \cline{2-5} 
    \multicolumn{1}{|c|}{}       & SD(${\langle \sigma \rangle}_{eq}$) & 1.11   & 1.43 & 0.92 \\ \cline{2-5} 
    \multicolumn{1}{|c|}{}         & SD(${\langle \sigma_{eq} \rangle}$) & 1.06  & 1.29 & 0.75 \\ \hline
    \end{tabular}
    \caption{\label{tab:avgs_stds_stress} Realization-wise averages and standard deviations of the yield
    stresses (in MPa) for each configuration of twinning and each load direction.}
    \end{table}

Due to the isotropy of the Voronoi tessellation and the fact that crystalline orientations were sampled from a uniform distribution on the space of
rotation matrices, the results are statistically independent of the direction of the prescribed load. In order to verify that, the responses to loads
along $x$, $y$ and $z$ axes were analyzed (Table~\ref{tab:avgs_stds_stress}).

Our results show that although slightly, structures without twins are more resistant than structures containing neighboring twins,
whereas polycrystals with inclusion twins provide the least resistant response. Additionally, the difference in dispersion regarding yield stresses
is partly explained by the difference in dispersion concerning average grain sizes (Table~\ref{tab:grain_nb_and_size}).
While the average yield stresses differ little, their high dispersion requires careful consideration. Statistical
significance tests should be appropriately performed to assess the representativity of these values. For that purpose,
Welch's $t$-test was chosen, due to it being applicable to populations with unequal variances. Additionally, it assumes
that both populations are normally distributed.
This condition is asymptotically satisfied when the domain is large, for certain diffusion problems with elliptic PDEs~\cite{nolen2014normal}.
The tested null hypothesis is that the two populations have equal mean, in which a two-tailed test is applied.
Table~\ref{tab:welch} shows that for $5\%$
significance level, the null hypothesis of equal means in only rejected for all scenarios when comparing inclusion twins and no twins populations.

\begin{table}[t]
    \resizebox{\textwidth}{!}{\begin{tabular}{|c|cccccc|cccccc|}
        \hline
        \multirow{3}{*}{} & \multicolumn{6}{c|}{Test for ${\langle \sigma \rangle}_{eq}$}                                                                                                                     & \multicolumn{6}{c|}{Test for $\overline{\langle \sigma_{eq} \rangle}$}                                                                                                                     \\ \cline{2-13} 
                          & \multicolumn{2}{c|}{x}                                  & \multicolumn{2}{c|}{y}          
                                                 & \multicolumn{2}{c|}{z}             & \multicolumn{2}{c|}{x}         
                                                                          & \multicolumn{2}{c|}{y}                                  & \multicolumn{2}{c|}{z}             \\ \cline{2-13} 
                          & \multicolumn{1}{c|}{p-value}  & \multicolumn{1}{c|}{NH} & \multicolumn{1}{c|}{p-value} 
                           & \multicolumn{1}{c|}{NH} & \multicolumn{1}{c|}{p-value}  & NH & \multicolumn{1}{c|}{p-value} 
                            & \multicolumn{1}{c|}{NH} & \multicolumn{1}{c|}{p-value}  & \multicolumn{1}{c|}{NH} & \multicolumn{1}{c|}{p-value}
                              & NH \\ \hline
        NeT vs InT        & \multicolumn{1}{c|}{1.08$\cdot10^{-1}$} & \multicolumn{1}{c|}{NR} & \multicolumn{1}{c|}{2.14$\cdot10^{-2}$}
         & \multicolumn{1}{c|}{R}  & \multicolumn{1}{c|}{4.31$\cdot10^{-3}$} & R  & \multicolumn{1}{c|}{3.68$\cdot10^{-2}$} & \multicolumn{1}{c|}{R} 
          & \multicolumn{1}{c|}{3.02$\cdot10^{-3}$} & \multicolumn{1}{c|}{R}  & \multicolumn{1}{c|}{9.77$\cdot10^{-4}$} & R  \\ \hline
        NeT vs NoT        & \multicolumn{1}{c|}{5.98$\cdot10^{-2}$} & \multicolumn{1}{c|}{NR} & \multicolumn{1}{c|}{1.11$\cdot10^{-1}$}
         & \multicolumn{1}{c|}{NR} & \multicolumn{1}{c|}{2.96$\cdot10^{-1}$} & NR & \multicolumn{1}{c|}{3.12$\cdot10^{-2}$} & \multicolumn{1}{c|}{R} 
          & \multicolumn{1}{c|}{8.70$\cdot10^{-2}$} & \multicolumn{1}{c|}{NR} & \multicolumn{1}{c|}{9.77$\cdot10^{-2}$} & NR \\ \hline
        InT vs NoT        & \multicolumn{1}{c|}{1.42$\cdot10^{-4}$} & \multicolumn{1}{c|}{R}  & \multicolumn{1}{c|}{4.62$\cdot10^{-4}$}
         & \multicolumn{1}{c|}{R}  & \multicolumn{1}{c|}{8.67$\cdot10^{-5}$} & R  & \multicolumn{1}{c|}{9.08$\cdot10^{-6}$} & \multicolumn{1}{c|}{R}
           & \multicolumn{1}{c|}{4.10$\cdot10^{-5}$} & \multicolumn{1}{c|}{R}  & \multicolumn{1}{c|}{3.37$\cdot10^{-7}$} & R  \\ \hline
        \end{tabular}}
        \caption{\label{tab:welch} Results of the statistical significance tests for both yield stress measures and all three load directions.
        Two-tailed Welch's $t$-test was performed for the null hypothesis of equal means. ``NeT'', ``InT'' and ``NoT'' denote respectively
        ``neighboring twins'', ``inclusion twins'' and ``no twins''. NH denotes ``null hypothesis'', while ``NR'' and ``R'' denote respectively
        ``not rejected'' and ``rejected''.}
        \end{table}    

\begin{figure}
\begin{center}
\pgfplotsset{width=0.7\linewidth, height=0.4\linewidth}
\begin{tikzpicture}
\begin{axis}[
title={},
xlabel={${\sigma}_{eq}$ (MPa)},
ylabel={Probability density},
legend style={at={(0.5,0)},anchor=south,nodes={scale=0.8, transform shape}},
xmin=100, xmax=250,
ymin=0, ymax=0.012,
]
\addplot+ [mark=o,mark size=0.5pt,color=blue] table [
            ] {fig6a.dat};
\addplot+ [mark=o,mark size=0.5pt,color=green] table [
            ] {fig6b.dat};
\addplot+ [mark=o,mark size=0.5pt,color=red] table [
            ] {fig6c.dat};
\legend{
No twins,
Neighboring twins, 
Inclusion twins,
}
\end{axis}
\end{tikzpicture}
\caption{Probability density function (PDF) of the von Mises stress field ${\sigma}_{\textnormal{eq}}$ estimated
for the set of 15 realizations for each twinning configuration subjected to a load along the $x$ axis
($251.7$ million voxels used for computing each curve).}
\label{fig:pdf_and_cdf}
\end{center}
\end{figure}
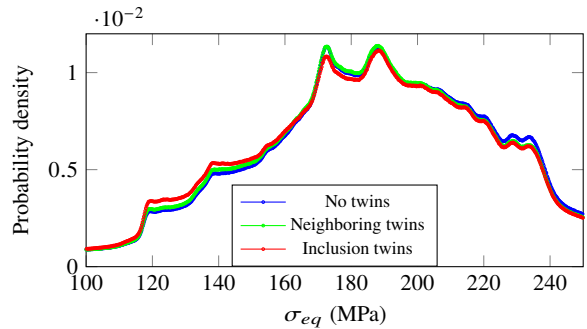

The probability density function of ${\sigma}_{eq}$ over all realizations is visualized in Fig.~\ref{fig:pdf_and_cdf}.
The graph shows that the distributions are multimodal. There are essentially two highest peaks likely associated to the initial
resistance to slip thresholds from Eq.~(\ref{eq:plastic_params}).
In order to interpret tehse results, the averages of twin volume fraction $f_{\beta}$ and accumulated slip $v^s$ fields were estimated
for each twinning configuration (Table~\ref{tab:fb_nu}), by averaging over slip systems, voxels and realizations. On the one hand, the averages for
neighboring twins and no twins are the same. On the other hand, RVEs with inclusion twins present systematically higher averages for
$\langle f_{\beta} \rangle$ and lower averages for $\langle v^s \rangle$. While the absence of difference between neighboring twins and
no twins confirms the lack of influence of crystallographic twins on the macroscopic elastoplastic response, the values obtained for inclusion
twins hints at an interaction between morphology and texture.

\begin{table}[t]
    \centering
    \begin{tabular}{|cc|c|c|c|}
    \hline
    \multicolumn{2}{|c|}{}        & Load along x & Load along y & Load along z \\ \hline
    \multicolumn{1}{|c|}{\multirow{4}{*}{Neighboring twins}} & $\overline{\langle f_{\beta} \rangle}$
      & 2.24$\cdot10^{-4}$     & 2.26$\cdot10^{-4}$     & 2.26$\cdot10^{-4}$     \\ \cline{2-5} 
    \multicolumn{1}{|c|}{}     & $\overline{\langle v^s \rangle}$   & 1.55$\cdot10^{-4}$     & 1.55$\cdot10^{-4}$ 
        & 1.55$\cdot10^{-4}$     \\ \cline{2-5} 
    \multicolumn{1}{|c|}{}      & SD($\langle f_{\beta} \rangle$) & 6.27$\cdot10^{-6}$     & 5.73$\cdot10^{-6}$  
       & 3.60$\cdot10^{-6}$     \\ \cline{2-5} 
    \multicolumn{1}{|c|}{}       & SD($\langle v^s \rangle$) & 2.06$\cdot10^{-6}$     & 2.60$\cdot10^{-6}$ 
        & 1.44$\cdot10^{-6}$     \\ \hline
    \multicolumn{1}{|c|}{\multirow{4}{*}{Inclusion twins}} & $\overline{\langle f_{\beta} \rangle}$  
     & 2.41$\cdot10^{-4}$     & 2.42$\cdot10^{-4}$     & 2.41$\cdot10^{-4}$     \\ \cline{2-5} 
    \multicolumn{1}{|c|}{}       & $\overline{\langle v^s \rangle}$   & 1.51$\cdot10^{-4}$  
       & 1.51$\cdot10^{-4}$     & 1.52$\cdot10^{-4}$     \\ \cline{2-5} 
    \multicolumn{1}{|c|}{}       & SD($\langle f_{\beta} \rangle$) & 4.56$\cdot10^{-6}$  
       & 5.17$\cdot10^{-6}$     & 5.11$\cdot10^{-6}$     \\ \cline{2-5} 
    \multicolumn{1}{|c|}{}     & SD($\langle v^s \rangle$) & 1.28$\cdot10^{-6}$   
      & 2.18$\cdot10^{-6}$     & 1.73$\cdot10^{-6}$     \\ \hline
    \multicolumn{1}{|c|}{\multirow{4}{*}{No twins}} & $\overline{\langle f_{\beta} \rangle}$  
     & 2.24$\cdot10^{-4}$     & 2.24$\cdot10^{-4}$     & 2.26$\cdot10^{-4}$     \\ \cline{2-5} 
    \multicolumn{1}{|c|}{}     & $\overline{\langle v^s \rangle}$   & 1.56$\cdot10^{-4}$ 
        & 1.56$\cdot10^{-4}$     & 1.55$\cdot10^{-4}$     \\ \cline{2-5} 
    \multicolumn{1}{|c|}{}     & SD($\langle f_{\beta} \rangle$) & 7.13$\cdot10^{-6}$   
      & 8.56$\cdot10^{-6}$     & 6.68$\cdot10^{-6}$     \\ \cline{2-5} 
    \multicolumn{1}{|c|}{}       & SD($\langle v^s \rangle$) & 2.37$\cdot10^{-6}$   
      & 3.13$\cdot10^{-6}$     & 2.18$\cdot10^{-6}$     \\ \hline
    \end{tabular}
    \caption{\label{tab:fb_nu}
    Realization-wise averages and standard deviations of the slip-system-wise and voxel-wise averages of the elastoplastic variables
    $f_\beta$ and $v^s$ (dimensionless) for each configuration of twinning and each load direction.}
    \end{table}

A final remark concerns the Hall-Petch effect,
which states that the yield strength of a polycrystalline materials decreases as $d^{-1/2}$,
where $d$ is the average grain size,
for values of $d$ larger than a certain critical size.
Indeed, materials with larger grains in average tends to present a higher number of dislocations
piling up at grain boundaries, favoring the transmission of dislocations to other grains.
When the grain size becomes smaller than the critical size, on the contrary,
experimental observations~\cite{conrad_grain_2000} have shown that 
the yield strength actually decreases when $d$ decreases.
In this range of grain refinement, dislocation slip dominates other mechanisms of plastic flow,
such as grain boundary sliding.
In any case, we emphasize that the Hall-Petch relationship,
while being a significant phenomenon to consider,  is not taken into account
in the present work.

\section{Conclusion}\label{sec:conclusion}
The contribution of this study is twofold. First, we proposed a regularization strategy for 
accelerating an explicit integration scheme of a rate-independent crystal plasticity law implemented on an FFT-based
solver. Second, the influence of twin grains with different morphologies on the macroscopic yield stress has been investigated,
making use of stochastically-generated polycrystalline RVEs of a $\gamma$-TiAl intermetallics material with tetragonal crystal symmetry.
Our result indicates that the presence of twins has no significant effect on
the average elastoplastic behavior of polycrystals, provided untextured crystallographic orientations are considered.
However, the interaction of grain morphology and size with local correlations arising from twin relations has to be further analyzed.

\begin{acknowledgement}
We are indebted to S. Forest for useful suggestions made on a previous version of this paper.
The authors acknowledge the financial support of the French Agence Nationale de la Recherche (ANR, ANR-21-FAI1-0003)
and the Bundesministerium f\"ur Bildung und Forschung (BMBF, 01IS2109) within the French-German research project
SMILE\footnote{See \url{https://people.cmm.minesparis.psl.eu/users/willot/Smile/}.}.
\end{acknowledgement}

\bibliographystyle{elsarticle-num}
\bibliography{MonteiroEtAl2024cmds}

\end{document}